\documentclass[reprint,amsmath,amssymb,aip,cha]{revtex4-2}

\usepackage{graphicx}
\usepackage[utf8]{inputenc}
\usepackage[T1]{fontenc}
\usepackage{newtxtext}
\usepackage{newtxmath}
\usepackage{microtype}
\usepackage{siunitx}
\usepackage{physics}
\usepackage[svgnames]{xcolor}
    \definecolor{accent}{HTML}{df2d16}
\usepackage[allcolors=accent,colorlinks,pdfusetitle,pdfauthor={Níckolas de Aguiar Alves}]{hyperref}
\usepackage{hyperref}
\usepackage{orcidlink}

\begin{document}
\title{The unexpected effectiveness of the Picard maneuver: Seeing three images of the same superluminal spaceship}
\thanks{The following article has been accepted by the \href{https://pubs.aip.org/aapt/ajp}{\emph{American Journal of Physics}}. After it is published, it will be found at \url{https://doi.org/10.1119/5.0353857}.}

\author{\firstname{Níckolas} de \surname{Aguiar Alves}\,\orcidlink{0000-0002-0309-735X}}
\email{alves.nickolas@ufabc.edu.br}
\affiliation{Center for Natural and Human Sciences, \href{https://ror.org/028kg9j04}{Federal University of ABC}, Avenida dos Estados 5001, Bangú, Santo André, São Paulo 09280-560, Brazil}

\begin{abstract}
    The USS Stargazer, commanded by Captain Jean-Luc Picard, battles a Ferengi vessel. In an effort to survive the battle, the Stargazer does the unthinkable: they move toward the Ferengi vessel at a faster-than-light speed, stopping just before a crash. The Ferengi vessel is said to see two images of the Stargazer, and ends up shooting the wrong one. In this note, I show with spacetime diagrams that, in fact, the Ferengi vessel sees up to three images of the Stargazer. This gives a simple example of how there can be multiple values of retarded time when an object moves faster than light, and how these multiple values are not restricted to only two when the object can accelerate.
\end{abstract}

\maketitle

\begin{quote}
    ``With the enemy vessel coming in for the kill, I ordered a sensor bearing, and when it came into the return arc---'' \\
    ``You performed what Starfleet textbooks now refer to as the `Picard maneuver.''' \\
    ``Well, I did what any good helmsman would've done---I dropped into high warp, stopped right off the enemy vessel's bow and fired with everything I had.'' \\ 
    ``And blowing into maximum warp speed, you appeared, for an instant, to be in two places at once!''\\
    ``And our attacker fired on the wrong one!'' \cite{bowman1987TheBattle}
\end{quote}

When I first watched ``The Battle,''\cite{bowman1987TheBattle} I was already acquainted with special relativity, and hearing the conversation above was delightful. It is a beautiful exploration, within science fiction, of the concept of retarded time and of how Captain Jean-Luc Picard managed to exploit it. 

Let us start with the basic ideas. Because light travels at a finite speed \(c\), an observer sitting at position \(\vb{r}\), at time \(t\), sees a spaceship at the position \(\vb{X}(t_{\text{ret}})\) it had at retarded time \(t_{\text{ret}}\). Retarded time is given implicitly by the equation\cite{griffiths2023IntroductionElectrodynamics} 
\begin{equation}\label{eq: tret-definition}
    \norm{\vb{r} - \vb{X}(t_{\text{ret}})} = c (t - t_{\text{ret}}).
\end{equation}
Note that, if we fix \(t\) and \(\vb{r}\), the number of images seen by the observer is the number of solutions of Eq. \eqref{eq: tret-definition}. If the spaceship is moving slower than the speed of light, there can be at most one solution. A simple argument is given by Griffiths, in Sec. 10.3.1 of Ref. \onlinecite{griffiths2023IntroductionElectrodynamics}. Here is another version. Suppose there were two different solutions \(t_1\) and \(t_2\) for Eq. \eqref{eq: tret-definition}. We label them so \(t_2 > t_1\). Then 
\begin{equation}\label{eq: two-solutions-tret}
    \frac{\norm{\vb{r} - \vb{X}(t_1)} - \norm{\vb{r} - \vb{X}(t_2)}}{t_2 - t_1} = c.
\end{equation}
To proceed, let us recall the reverse triangle inequality: for any two vectors \(\vb{u}\) and \(\vb{v}\), it holds that\cite{axler2024LinearAlgebraDone} 
\begin{equation}
    \big|\norm{\vb{u}}-\norm{\vb{v}}\big| \leq \norm{\vb{u}-\vb{v}}.
\end{equation}
If we use the reverse triangle inequality on the left-hand side of Eq. \eqref{eq: two-solutions-tret}, then we conclude that
\begin{equation}\label{eq: two-images-cant-be-subluminal}
    \frac{\norm{\vb{X}(t_1) - \vb{X}(t_2)}}{t_2 - t_1} \geq c.
\end{equation}
This means the average speed of the spaceship would have been at least the speed of light! Therefore, in subluminal motion, there can be at most one value of retarded time.

\begin{figure}[tb]
    \centering
    \includegraphics{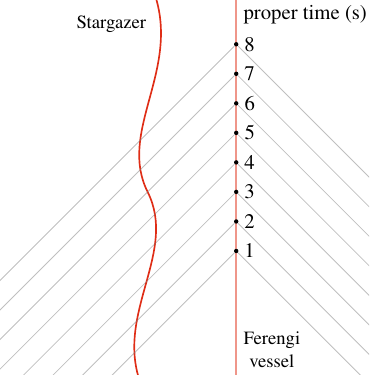}
    \caption{If the Stargazer is moving at subluminal velocities, then the Ferengi vessel will see at most one image of our heroic spaceship. In this diagram, the Ferengi see \emph{precisely} one image of the Stargazer at each instant.}
    \label{fig: subluminal}
\end{figure}

Let us picture this in a spacetime diagram. For concreteness, let me also borrow the terminology from ``The Battle.''\cite{bowman1987TheBattle} Our heroic spaceship is the USS Stargazer, and it is engaged in a battle against a Ferengi vessel.\cite{ferengi} In all diagrams, I will keep the Ferengi vessel still, while the Stargazer moves around. Hence, the diagrams are drawn in the Ferengi reference frame. With this background, we can take a look at Fig. \ref{fig: subluminal}, which illustrates the Stargazer moving at subluminal speeds. It intersects each past lightcone of the Ferengi vessel precisely once, and thus the vessel always sees a single image of the Stargazer at each time.\cite{acceleration}

The key concept explored in the Picard maneuver is that, if the Stargazer can move faster than light, then Eq. \eqref{eq: tret-definition} admits multiple solutions. I will not dive into what would take for a spaceship to move faster than light---for that, see Refs. \onlinecite{alcubierre1994WarpDrive,shoshany2019LecturesFasterThanLight}. To avoid extra complications, I will ignore the additional warping of spacetime that could be involved in the process. Hence, in the following, we will simply imagine that somehow it is possible for a spaceship to move faster than light in an approximately flat spacetime. Alternatively, you can imagine we are talking about fast particles moving through material media.\cite{material}

\begin{figure}[tb]
    \centering
    \includegraphics{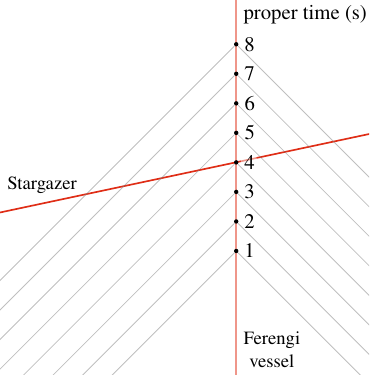}
    \caption{If the Stargazer is moving at a constant superluminal velocity, then the Ferengi vessel will either see no images of the Stargazer, or it will see two. There is one single instant (\(t = \SI{4}{\second}\)) in which the vessel sees a single image of the Stargazer.}
    \label{fig: superluminal}
\end{figure}

Let us then imagine that the Stargazer moves at a constant velocity \(\vb{v}=v\vu{z}\), with \(v > c\). Then Eq. \eqref{eq: tret-definition} can be rewritten as 
\begin{equation}
    c (t - t_{\text{ret}}) = \sqrt{x^2 + y^2 + (z - v t_{\text{ret}})^2}.
\end{equation}
If we square both sides of this expression, we find a quadratic equation for \(t_{\text{ret}}\). Hence, there are \emph{at most} two solutions. In practice, for a constant superluminal velocity, there are either two solutions, or there are zero solutions. We can have a single solution in special cases in which the two roots coincide. This can be seen algebraically (see, e.g., Sec. 23.7 in Ref. \onlinecite{zangwill2013ModernElectrodynamics}), but we will discuss it with a spacetime diagram---namely, the one on Fig. \ref{fig: superluminal}. For \(t < \SI{4}{\second}\), the worldline of the Stargazer has not yet crossed the Ferengi lightcone, meaning the spaceship is still too far away for the Ferengi to see. When \(t = \SI{4}{\second}\), the Ferengi see a single image of the Stargazer. For \(t > \SI{4}{\second}\), however, the worldline crosses the lightcones in two different points. Hence, the Ferengi would see the Stargazer at two different places. This is one way of understanding the quotation at the beginning of the paper. If a spaceship moves faster than light, its enemies may see it at two places at once.

\begin{figure}[tb]
    \centering
    \includegraphics{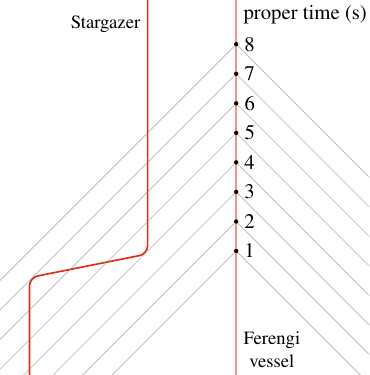}
    \caption{The Picard maneuver! The Stargazer begins at rest with respect to the Ferengi vessel. It rapidly accelerates to beyond the speed of light, and then stops soon after. Due to the acceleration periods, the Ferengi vessel may see not only two images of the Stargazer, but at times even three!}
    \label{fig: picard}
\end{figure}

Nevertheless, this is not the end of the story! In the actual Picard maneuver described in the quotation, the Stargazer starts at subluminal speeds, accelerates to a superluminal speed, and then decelerates back to subluminal speeds. Therefore, the inertial picture in Fig. \ref{fig: superluminal} is not accurate. For simplicity, let us assume the Stargazer and the Ferengi vessel start at rest relative to each other, and end at rest relative to each other. (Subluminal velocities would complicate the diagrams, but not make any significant changes to the argument.) With this in mind, the correct spacetime diagram for the Picard maneuver is, in fact, closer to the one on Fig. \ref{fig: picard}. The interested reader will find a more detailed mathematical model in App. \ref{app: math-model}, but we will not need it in the following discussion.

Figure \ref{fig: picard} reveals an extremely interesting property: at certain instants, the Ferengi vessel sees up to three images of the Stargazer! Let us keep track of what happens at each instant.
\begin{itemize}
    \item At early times (\(t < \SI{4}{\second}\)), the Ferengi see a single image of the Stargazer.
    \item Roughly at \(t = \SI{4}{\second}\), a second image of the Stargazer appears. This is the image of the spaceship after the maneuver. 
    \item For \(\SI{4}{\second} < t < \SI{7}{\second}\), the Ferengi vessel sees \emph{three} images of the Stargazer. There is one image from before the Picard maneuver began, one image from after the maneuver ended, and one image of the maneuver itself. As Ferengi proper time progresses, the image of the maneuver is seen as running backwards. 
    \item Roughly at \(t = \SI{7}{\second}\), the image of the maneuver is seen merging with the image from before the maneuver. Momentarily, there are only two images. 
    \item At late times (\(t > \SI{7}{\second}\)), the Ferengi once again see a single image of the Stargazer.
\end{itemize}

\begin{figure}[tb]
    \centering
    \includegraphics{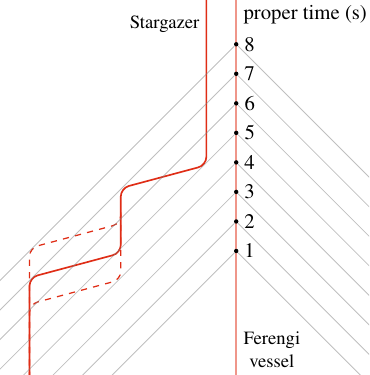}
    \caption{A multistep Picard maneuver. The Stargazer begins at rest with respect to the Ferengi vessel. It rapidly accelerates to beyond the speed of light, and then stops soon after. Rather than staying still, the spaceship does the same maneuver again. In this example, the Ferengi vessel can see up to five images of the Stargazer. Notice it is also possible to manipulate the order in which the images appear: the ``middle image'' can be made to appear before or after the ``current image,'' as illustrated by following the dashed paths.}
    \label{fig: picard-multi}
\end{figure}

In ``The Battle,''\cite{bowman1987TheBattle} the image of the Stargazer during the maneuver itself was ignored. This led to the incorrect depiction of only two images of the spaceship, when in fact the Ferengi vessel would have seen three.\cite{blur} A good enough helmsman could also further deceive their opponent by ``composing'' the Picard maneuver---for sufficiently complicated trajectories, even more copies of the spaceship would be seen by the Ferengi. One such example is given on Fig. \ref{fig: picard-multi}.

Particles can move faster than light in material media. If they accelerate, there may be more than two solutions to Eq. \eqref{eq: tret-definition}. From a mathematical perspective, this is understood in that the function \(\vb{X}(t_{\text{ret}})\) can be extremely complicated, and therefore Eq. \eqref{eq: tret-definition} may not reduce to a quadratic equation. The Picard maneuver illustrates this phenomenon in a simple way, which can be understood through the use of spacetime diagrams. 

\begin{acknowledgments}
    I thank two anonymous reviewers for valuable suggestions---in particular, the use of the reverse triangle inequality to obtain Eq. \eqref{eq: two-images-cant-be-subluminal}.

    The work of NAA was supported by the São Paulo Research Foundation (FAPESP) under grant 2025/05161-0.
\end{acknowledgments}

\appendix
\section{Mathematical Model for the Picard Maneuver}\label{app: math-model}
    While much about the Picard maneuver can be understood in terms of spacetime diagrams, it can be useful to have a more detailed mathematical model---for example, to highlight what Eq. \eqref{eq: tret-definition} looks like. I suggest the following. The trajectory of the Stargazer can be modelled as
    \begin{equation}\label{eq: mathematical-picard-model}
        x(t) = x_i + (x_f - x_i) H\qty(\frac{v t}{x_f - x_i}),
    \end{equation}
    where \(x_i\) is the initial position, \(x_f\) is the final position, and \(v\) is the velocity at time \(t = 0\). \(H(\lambda)\) is a function modelling the acceleration phase. It is convenient to assume it has the properties
    \begin{subequations}
        \begin{gather}
            \lim_{\lambda \to + \infty} H(\lambda) = 1, \\
            \lim_{\lambda \to - \infty} H(\lambda) = 0, \\
            H'(0) = 1.
        \end{gather}
    \end{subequations}
    This model leads to 
    \begin{equation}
        \dot{x}(t) = v H'\qty(\frac{v t}{x_f - x_i}),
    \end{equation}
    where the dot denotes a derivative with respect to time \(t\), and the prime denotes a derivative with respect to \(\lambda = vt/(x_f - x_i)\). The properties of the \(H\) function then ensure that \(\dot{x}(0) = v\). If \(H\) is sufficiently well-behaved, then \(\dot{x}(t)\) vanishes at early and late times (\(t \to - \infty\) and \(t \to +\infty\), respectively).
    
    Figure \ref{fig: picard} is based on a piecewise choice of \(H(\lambda)\), up to the rounded edges. A possible choice is then to pick 
    \begin{equation}\label{eq: piecewise-H}
        H(\lambda) = \begin{cases}
            0, & \text{for } \lambda \leq - \frac{1}{2}, \\
            \lambda + \frac{1}{2}, & \text{for } - \frac{1}{2} < \lambda < \frac{1}{2}, \\
            1, & \text{for } \frac{1}{2} \leq \lambda.
        \end{cases}
    \end{equation}
    If a smooth \(H(\lambda)\) is preferred, then one can use 
    \begin{equation}\label{eq: tanh-H}
        H(\lambda) = \frac{\tanh(2 \lambda) + 1}{2}.
    \end{equation}
    A third option is to consider the smooth, but non-holomorphic, function 
    \begin{equation}\label{eq: nonholomorphic-H}
        \theta(\lambda) = \begin{cases}
            0, & \text{for } \lambda \leq 0, \\
            \exp(-\frac{1}{\lambda}), & \text{for } \lambda > 0,
        \end{cases}
    \end{equation}
    and then define 
    \begin{equation}
        H(\lambda) = \frac{\theta\qty(\lambda + \frac{\sqrt{2}}{2})}{\theta\qty(\lambda + \frac{\sqrt{2}}{2}) + \theta\qty(\frac{\sqrt{2}}{2} - \lambda)}.
    \end{equation}
    This is a standard example of a smooth transition function, with coefficients chosen to ensure \(H'(0) = 1\).

    Notice that, depending on the choice of \(H(\lambda)\), \(v\) may not be the maximum velocity achieved during the trajectory. For example, the choice given in Eq. \eqref{eq: nonholomorphic-H} does not have maximum velocity at \(t = 0\), and thus the choice \(v=c\) would still include superluminal speeds. Equations \eqref{eq: piecewise-H} and \eqref{eq: tanh-H} lead to maximum speed at \(t=0\), and thus the velocity is capped by \(v\) in those cases. 

    Using these examples, one can see how varying the value of \(v\) changes the number of solutions to Eq. \eqref{eq: tret-definition}. This can be done, for instance, using graphical methods to solve the transcendental algebraic equations.


\begin{thebibliography}{99}
\bibitem{acceleration} If the Stargazer is moving at subluminal velocities, it \emph{is} possible that it cannot be seen by the Ferengi vessel. This can happen if the spaceship is moving with constant proper acceleration---see, e.g., Prob. 10.18 in Ref. \onlinecite{griffiths2023IntroductionElectrodynamics}.
\bibitem{material} Recall that the speed of light in a medium with refractive index \(n\) is \(c_n = c/n\). For \(n > 1\), we can have particles with \(c_n < v < c\). See, for example, Sec. 23.7 in Ref. \onlinecite{zangwill2013ModernElectrodynamics}. The discussion in this paper would still hold in this case, but one should replace all instances of \(c\) outside this footnote with \(c_n\).
\end{thebibliography}
\end{document}